# Presence of $^{236}U$ and $^{237}Np$ in a marine ecosystem: the northern Benguela Upwelling System, a case study

*Mercedes López-Lora[1,2], Elena Chamizo[1], Martina Rožmarić[3], Deon C. Louw[4]*

[1]Centro Nacional de Aceleradores (CNA). Universidad de Sevilla. Junta de Andalucía. Consejo Superior de Investigaciones Científicas. Parque científico y tecnológico Cartuja. Thomas Alva Edison 7, 41092, Sevilla, Spain

[2]Dpto. de Física Aplicada I, Escuela Politécnica Superior, Universidad de Sevilla, Virgen de África 7, 41011 Sevilla, Spain

[3]International Atomic Energy Agency, Environment Laboratories, MC 98000 Monaco

[4]National Marine Information and Research Centre (NatMIRC), PO Box 912, Swakopmund, Namibia

*Corresponding author:
E-mail address: mlopezlora@us.es

## Abstract

The Benguela Upwelling System (BUS), off the south-western African coast, is one of the four major eastern boundary upwelling ecosystems in the oceans. However, despite its very interesting characteristics, this area has been almost overlooked in the field of environmental radioactivity. In this work, it has been carried out for the first time the combined study of $^{236}U$ and $^{237}Np$ in the coast of Namibia within the northern BUS. Surface seawater exhibited similar $^{236}U$ and $^{237}Np$ concentrations, ranging from $3.9 \cdot 10^6$ to $5.6 \cdot 10^6$ atoms kg$^{-1}$ and from $4.6 \cdot 10^6$ to $8.5 \cdot 10^6$ atoms kg$^{-1}$, respectively. The observed inventories in a water column from the continental margin, of $(2.10 \pm 0.11) \cdot 10^{12}$ atoms m$^{-2}$ for $^{236}U$ and $(3.48 \pm 0.13) \cdot 10^{12}$ atoms m$^{-2}$ for $^{237}Np$, were in agreement with the global fallout (GF) source term in the Southern Hemisphere, which was recognized as the main source of actinides to the region. A pattern was observed in the surface samples, with $^{237}Np$ concentrations that decreased by 25-30% when moving from inshore to offshore stations, but such an effect could not be clearly discerned in the case of $^{236}U$ within the data uncertainties. An explanation based on the larger particle reactivity of GF $^{237}Np$ compared to GF $^{236}U$ was proposed. Such an effect would have been important at the studied site due to the enhance presence of particles in the continental shelf triggered by the upwelling phenomenon. A value of $1.77 \pm 0.20$ was obtained for the $^{237}Np/^{236}U$ atom ratio for the GF source term in the marine environment.

## 1. Introduction

Since the massive release of anthropogenic radioactivity to the general environment in the 1960s, anthropogenic radionuclides have demonstrated to be key tools in the study of marine processes. $^{236}$U ($T_{1/2}$=23.4 My) belongs to the group of conservative oceanic tracers. Widely studied in the last decade mostly thanks to the achievements obtained in Accelerator Mass Spectrometry (AMS), $^{236}$U has been targeted in a multiplicity of contexts: i) in the North Atlantic Ocean, surrounding seas and Arctic Ocean, to track the signal from the effluents of European Nuclear Fuel Reprocessing plants (NRPs) and get insights into transient times (Casacuberta et al., 2018, 2016, 2014) ; ii) in the Mediterranean and Japan seas, to assess its levels and get information on the source term (Castrillejo et al., 2014; E Chamizo et al., 2016; Sakaguchi et al., 2012); iii) in corals from the Pacific ocean, to establish the temporal trend of the fallout signal generated during the atmospheric testing of nuclear weapons (Winkler et al., 2012); and iii) in the North Pacific and Equatorial Pacific ocean, to evaluate the fallout signal and explore its potential to trace water masses carrying an extremely low anthropogenic $^{236}$U signal (Eigl et al., 2017; Villa-Alfageme et al., 2019). These studies are based on the basic assumption that $^{236}$U behaves as the naturally-occurring uranium isotopes in the oceans: mostly presented in oxidation state (VI) in well-oxygenated waters, forming stable complexes with carbonates and therefore behaving conservatively. However, the geochemical behaviour of anthropogenic $^{236}$U has not been thoroughly studied and in the case of the different $^{236}$U species associated to the effluents from NRPs this kind of studies might be especially relevant.

On the other hand, the situation of $^{237}$Np ($T_{1/2}$= 2.14 My) has been radically different. Having a long half-life and being a decay-product of $^{241}$Am ($T_{1/2}$= 432.2 y), the first reported studies were motivated by a radiological concern. Thus, some scientific reports dealing with its physico-chemical behaviour in the marine environment were published (Germain et al., 1987; McCubbin and Leonard, 1997). These works were focused on the speciation of Np, as well as on its transfer from seawater to sediments and marine organisms. The main conclusions drawn from these studies are: i) the most important Np specie in open seawater is the neptunyl ion $Np0_2^+$ (i.e. Np(V)), but soluble anionic and neutral complexes and particulate forms can be also detected; ii) Np(V) is highly soluble, but in the presence of sedimentary material can be reduced to the less soluble tetravalent form Np(IV), which shows slow oxidation kinetics and might persist in natural waters; iii) Np has a smaller probability of transfer to sediments and organisms than particle reactive radionuclides such as plutonium and americium. Moreover, it was found that approximately 50% of the $^{237}$Np from Sellafield NRP was released in its reduced form, being stored in the sediments, surviving between 1-0.5% of it in the surrounding seas (Hallstadius et al., 1986). However, a long-term assessment of the behaviour of Np in the general marine environment has not been carried out. On the other hand, very scarce data on the presence of Np in the general marine environment exist, with most of the studies focused on regions directly impacted by nuclear reprocessing plants (Beasley et al., 1998; Lindahl et al., 2005; Pentreath and Harvey, 1981). The reason for that are constraints of analytical nature due to i) the necessity of using large volume samples and a complex chemistry to analyse it by alpha-spectrometry or conventional mass spectrometry techniques, and ii) the lack of a long-life isotopic tracer (Thakur and Mulholland, 2012). In contrast, the recently reported $^{236}U$ studies have covered a wide part of the Northern Hemisphere, but the Southern Hemisphere is still almost neglected.

In this work, for the first time, the presence of $^{236}U$ and $^{237}Np$ in the Southern Atlantic Ocean was evaluated. The target area was the northern Benguela Upwelling System (nBUS), which is considered as the most important eastern boundary upwelling ecosystems. Upwelling is the phenomenon occurring in areas where winds blow parallel to the coastline, resulting in the transport of nutrient depleted surface waters away from the land, which are replaced by cold, bottom nutrient-rich water, which triggers dramatically primary production with a follow up on zooplankton and fish production. Subsequently,  upwelled water in the BUS feature seasonally a high content of organic particles, opal and carbonates, in addition to crustal material (Vorrath et al., 2018). Offshore deeper water masses are of Antarctic and Atlantic origin, with different particle composition. Therefore, the assessment of $^{236}U$ and $^{237}Np$ in such complex ecosystem might offer a unique opportunity to get information on the particle reactivity of both radionuclides and, consequently, on the potential of their use in a dual tracer approach in other ecosystems. As for general environmental studies, first results on their concentrations in an almost neglected region in the Southern Hemisphere can be gained and, from them, insights into their source terms could be obtained. It is important to emphasize that the only operational nuclear power plant in the entire African continent is in South Africa (the Koeberg nuclear power plant, which is approximately 700 km from the southern Namibian border), and no study has been reported regarding its impact on neighbouring seas. State-of-the-art AMS techniques set up at the CNA and a radiochemical procedure recently developed in the frame of the existing collaboration between the IAEA and the CNA, have been applied to achieve the previously explained goals (Chamizo and López-Lora, 2019; López-Lora et al., 2019; López-Lora and Chamizo, 2019).

## 2. Sources of anthropogenic $^{236}U$ and $^{237}Np$ to the South Atlantic Ocean

Based on the well-established latitudinal deposition dependency of the stratospheric fallout (i.e. global fallout, GF) (Hardy et al., 1973), about 1/3 of the overall budgets would have been deposited at negative latitudes (i.e. about 300 kg of $^{236}U$ and 500 kg of $^{237}Np$), becoming the most significant source of anthropogenic radioactivity in the South Atlantic Ocean. On the other hand, different studies have demonstrated that the Southern Hemisphere is strongly influenced by the tropospheric fallout from the low yield atmospheric nuclear tests performed in the French Polynesia (1966-1974) and in the Australian territory (1952-1958) by France and United Kingdom, respectively (Froehlich et al., 2019). Indeed, the $^{240}Pu/^{239}Pu$ atomic ratios in soils have revealed a much more heterogeneous pattern in the South Hemisphere than in the North one, probably because of the uneven distribution of debris coming from those tests (Kelley et al., 1999). The extent to which those tests could have affected the $^{236}U$ and $^{237}Np$ baseline levels in specific areas of the Southern Hemisphere has not been documented to date.

On a local scale, different sources should be considered (Fig. 1). The South Atlantic Ocean hosted the launch of 3 USA nuclear devices in 1958. However, being detonated in the upper atmosphere, local or regional fallout are not expected from those USA tests (UNSCEAR, 2008). Another source is the SNAP-9A satellite accident in 1964, as a consequence of which 1 kg of the $^{238}$Pu fuel from its radiothermal generator was released in the upper atmosphere over Madagascar, peaking its deposition at those latitudes (UNSCEAR, 2008). Since the production of $^{238}$Pu is based on the neutron activation of $^{237}$Np, the damaged SNAP-9A unit may have contained traces of $^{237}$Np, but the actual amounts are unknown. Other potential local sources of anthropogenic radioactivity to the Eastern South Atlantic Ocean are related to the nuclear industry. South Africa hosts the only nuclear power plant within the African continent (Koeberg), which achieved commercial operation in 1984-1985 and is located 30 km north of Cape Town (UNSCEAR, 2008). Low and intermediate level nuclear wastes coming from this plant are stored at the waste disposal facility of Vaalputs, located at 500 km north of Koeberg and 100 km to the coastline (International Atomic Energy Agency., 1997). To the best knowledge of the authors, actinide releases to the Atlantic Ocean from these facilities have not been reported so far.

## 3. Materials and methods

### 3.1 Study area

The most relevant surface circulation currents in the South Atlantic Ocean, including the area of interest in this study, the Namibian coast, are visible in Fig. 1. This region is characterized by the presence of the Benguela Current which, together with the South Equatorial Current (SEC), Brazil Current (BC) and South Atlantic Current (SAC), constitutes the anticyclonic southern subtropical gyre, dominating the water mass circulation in the South Atlantic Ocean (Peterson and Stramma, 1991).

The Benguela current system flows northwards covering South Africa coast from Cape Town (34°S) to the southern part of Angola (~15°S), including the entire Namibian coast, and forming the already-mentioned BUS, one of the most important upwelling ecosystems in the world (i.e. together with California, Northwest Africa and Peru) (Carr and Kearns, 2003; Shannon, 1985). There is an unusually intense cell of upwelling at Lüderitz (26°S) which effectively divides it into two parts (Hutchings et al., 2009) (Fig. 1-b). The southern Benguela upwelling system thus extends as far northwards as Lüderitz, while the rest of the Namibian coast falls within the northern Benguela upwelling system. Namibian coast is a hyper-arid desert characterized by a windy climate. Its continental shelf is generally narrow and is one of the deepest in the world, with an average shelf edge depth of 350 m (Shannon, 1985).

### 3.2 Sampling

In May 2014, a sampling campaign along the Namibian coast onboard the *RV Mirabilis* was organized in collaboration between the National Marine Information and Research Centre (NatMIRC) in Namibia, and the IAEA Environment Laboratories (IAEA-NAEL) in Monaco. Seawater samples for analysis of radionuclides were collected north of Lüderitz within the nBUS, up to an offshore distance of 170 km in areas of the Namibian continental shelf (Fig. 1-b). Surface samples were collected by pumping water from the surface onto deck, while deep water sampling was carried out using a rosette on which the conventional Niskin bottles were attached coupled to a CTD instrument (i.e. to measure the conductivity, temperature and density). Nineteen 2-5 L aliquots were collected for the independent analysis of $^{237}Np$ and $^{236}U$ by AMS, including a seawater profile for the deepest station (i.e. 1272 m depth, Station 7). They were directly sent to the CNA without further pre-treatment onboard (i.e. neither filtrated nor acidified) to avoid contamination problems from the reagents or other materials because of the very small concentration expected in these samples. For certain stations (i.e. 1, 2, 8, 9 and 11), additional aliquots from the IAEA repository (i.e. a surplus of the samples collected during the same campaign for the study of other elements) were used later on for validation purposes. The second aliquots had been acidified onboard, in contrast to the first batch, which were acidified in the laboratory prior to analysis.

### 3.3 Radiochemical separation of uranium and neptunium

The collected seawater samples were processed following the radiochemical procedure described in (López-Lora et al., 2019), based on the sequential separation of U and Np+Pu from the same seawater aliquot using a non-isotopic approach in the case of $^{237}Np$ (i.e. $^{242}Pu$ is used as yield tracer for both Pu and Np). Briefly, following the acidification and the spike addition (i.e. $^{233}U$ and $^{242}Pu$), U, Np and Pu isotopes were pre-concentrated from the seawater samples with $Fe(OH)_2$. After a proper adjustment of the oxidation states, U and Np + Pu combined fractions were sequentially eluted using UTEVA® and TEVA® resins placed in tandem, respectively, and subsequently adapted to the optimal matrix for the AMS analysis. The methodology described in (López-Lora et al., 2019) to keep at minimum the $^{237}Np$ and $^{236}U$ contamination was followed in this work to be able to analyse the extremely low $^{236}U$ and $^{237}Np$ concentrations that were expected in the samples. Different procedural blanks (i.e. 5L MQ® water aliquot spiked with $^{233}U$ and $^{242}Pu$) were processed together with the samples of interest to control the laboratory background levels. The QA/QC of the radiochemical procedure had been carried out using IAEA certified reference material IAEA-443 (Irish seawater) (López-Lora et al., 2019).

### 3.4 AMS measurements

$^{236}U$, $^{238}U$ and $^{237}Np$ AMS determinations were carried out with the compact 1 MV AMS system at the CNA, following the techniques described in (Chamizo and López-Lora, 2019) and (López-Lora and Chamizo, 2019). Briefly, i) U or Np species were injected into the accelerator as negative monoxide anions (i.e. $UO^-$, $NpO^-$ or $PuO^-$), iii) stripped to 3+ at the terminal of a tandem electrostatic accelerator working at about 670 kV in He gas (e.g. $^{237}Np^{16}O^- \rightarrow Np^{3+}$); iii) analysed on their E/q and M/q properties using a sector magnet and an electrostatic deflector, and iv) counted: a) in a gas ionization chamber in the case of the rare isotopes (i.e. $^{236}U$ and $^{233}U$ during an U measurements, $^{237}Np$ and $^{242}Pu$ during the Np one) , and b) in an offset Faraday Cup in the case of $^{238}U$ during an U measurement. The stripping process allows the removal of molecular isobars (e.g. $^{235}U^1H$ in the case of $^{236}U$ or $^{235}U^1H_2$ in the case of $^{237}Np$) providing the necessary sensitivity. After a thorough adjustment of the setup parameters, $^{236}U/^{238}U$ and $^{237}Np/^{238}U$ background atom ratios

of $10^{-10}$ and $10^{-11}$, respectively, were achieved and corrected during these experiments. An uranium sample Vienna-kkU was also measured together with the samples of interest, giving a $^{236}U/^{238}U$ atom ratio of $(8.2 \pm 2.0)\cdot10^{-11}$, in agreement with the expected value (i.e. $(6.98 \pm 0.32)\cdot10^{-11}$ (Steier et al., 2008)). Seawater samples spiked with $^{237}Np$ and $^{242}Pu$, which were included in every sample batch, turned out a deviation in the $^{237}Np$ results of 5% associated to the non-isotopic approach (i.e. chemistry procedure and AMS technique) as explained in (López-Lora and Chamizo, 2019). As a result, $^{237}Np$ concentrations can be analysed with a minimum 5% uncertainty. All the uncertainties presented in this article are expressed with k=1 (68%) confidence level.

## 4. Results

The obtained $^{236}U$ and $^{237}Np$ results for the seawater samples analysed in this work, including the ones corresponding to the IAEA repository, are summarized in Table S1. $^{238}U$ results and further sample details including oceanographic parameters (temperature, salinity and dissolved oxygen) are also specified in Table S1. $^{238}U$ results are discussed in the Supplemental Information section. In general, there is a good agreement between the twin aliquots within the uncertainties for both $^{236}U$ and $^{237}Np$. $^{236}U$ and $^{237}Np$ uncertainties exceed 10% in some cases due to the small sample volumes and the uncertainty sources associated to the AMS technique itself, as explained in (Chamizo and López-Lora, 2019; López-Lora and Chamizo, 2019). Results from the three control samples are in agreement with the expected values and show a consistent reproducibility of 5% for $^{237}Np$ concentrations. The results for the two analysed aliquots of the IAEA-443 reference sample match the previous reported values for this sample. Results from designated 'control samples' and IAEA-443 aliquots are discussed elsewhere (López-Lora and Chamizo, 2019). The studied procedural blanks were below the detection limit of the technique for both radionuclides (i.e. $10^6$ atoms per sample), meaning that no further corrections than the AMS inherent ones were applied to the results.

Measurements of surface seawater samples show similar $^{236}U$ and $^{237}Np$ concentration values. $^{236}U$ concentrations range from $3.9\cdot10^6$ to $5.6\cdot10^6$ atoms kg$^{-1}$ (i.e. in the 1-2 fg kg$^{-1}$ or 4-5 nBq kg$^{-1}$ range) and $^{236}U/^{238}U$ atom ratios from $4.8\cdot10^{-10}$ to $6.7\cdot10^{-10}$. These values are lower than the reported data for the Northern Hemisphere, but comparable with those published for seawater samples collected close to the equator at both the Atlantic and Pacific Oceans. In particular, $^{236}U$ concentrations in surface seawater in the North Atlantic Ocean varies from $10^7$ atoms kg$^{-1}$ (i.e. $^{236}U/^{238}U$ atom ratios at the level of $10^{-9}$) at middle latitudes, to $6\cdot10^6$ atoms kg$^{-1}$ (i.e. $^{236}U/^{238}U$ ratios of $6\cdot10^{-10}$) close to the equator (Casacuberta et al., 2014). These last values are similar to the previous data from the Equatorial Pacific Ocean (10.5-15°S), with $^{236}U$ concentrations in the $(3.9\text{-}6.6)\cdot10^6$ atoms kg$^{-1}$ range (i.e. $^{236}U/^{238}U$ atom ratios in the $(4.8\text{-}6.8)\cdot10^{-10}$ range) (Villa-Alfageme et al., 2019). For $^{237}Np$, concentrations are in the $(4.6\text{-}8.5)\cdot10^6$ atom kg$^{-1}$ range (i.e. from 50 to 90 nBq kg$^{-1}$). As expected, these values are considerably lower than the reported data for surface seawater samples in the North Hemisphere. For instance, values from 0.1 to 1 mBq kg$^{-1}$ have been reported for the Irish Sea (Pentreath and Harvey, 1981); in the 0.2-1 µBq kg$^{-1}$ range for the Arctic Ocean (Beasley et al., 1998); and in the 0.1-0.2 µBq L$^{-1}$ for the Mediterranean Sea (Bressac et al., 2017; Castrillejo Iridoy, 2017). To the best knowledge of the authors, no information on the $^{237}Np$ in seawater from the South Hemisphere has been reported so far.

$^{236}U$ and $^{237}Np$ results obtained from the water column at Station 7 are shown in Fig. 2. Both radionuclides show the same depth profile, decreasing from the surface to the deepest layers by more than one order of magnitude. Very small concentrations, at the limit of the technique, were obtained from the samples collected at the deepest point. The obtained $^{237}Np/^{236}U$ atom ratios do not change significantly along the profile within the uncertainties.

# 5. Discussion

## 5.1 Oceanic distribution and sources of $^{236}U$ and $^{237}Np$ along the Namibian Coast

Fig. 3 shows the $^{236}U$ and $^{237}Np$ results in T-S (potential temperature vs salinity) plots. Surface samples correspond to Modified Upwelling Waters (MUW) and show the highest $^{236}U/^{238}U$ and $^{237}Np$ values. Below this surface layer, samples taken from 300 m and 600 m at Station 7 are identified as Central Waters (CW) and show $^{236}U/^{238}U$ ratios and $^{237}Np$ concentrations in the $(0.7\text{-}4)\cdot10^{-10}$ range and in the $(1\text{-}6)\cdot10^{6}$ atom kg$^{-1}$ range, respectively. Finally, the deepest sample studied in this work (i.e. collected at 1250 m at Station 7) might be influenced by Antarctic Intermediate Waters (AAIW) and North Atlantic Deep Waters (NADW). This sample shows the lowest $^{236}U/^{238}U$ atom ratio, at the $2\cdot10^{-11}$ level, and $^{237}Np$ concentrations below $10^{6}$ atoms kg$^{-1}$, the detection limit of the technique. Reported $^{236}U/^{238}U$ ratios for AAIW and NADW in the North Atlantic Ocean are higher than the obtained value for this bottom sample (Table 1) due to the dilution and mixing effects experienced by these water masses on their way to the studied site.

The $^{236}U$ and $^{237}Np$ full depth concentration profiles (Fig. 3) show the typical diffusion profiles of a soluble element with an exclusively input from the surface. This trend was expected, since both elements have mostly a conservative behaviour in seawater and no advective transport of radionuclides with water masses carrying a local signal could be anticipated at the studied site. Figure 2 compares the obtained $^{236}U/^{238}U$ profile with other two profiles from different areas: i) from the Equatorial Pacific Ocean, where GF is the only source to be considered for $^{236}U$ and $^{237}Np$ (Villa-Alfageme et al., 2019); and ii) from the North Atlantic Ocean, where the formation of deep water carrying the signal of European NRP influences the shape of the profile (Casacuberta et al., 2014). The profile from the Namibian coast is similar to the one from the Equatorial Pacific, matching not only its shape but also the $^{236}U/^{238}U$ values which might indicate a similar input source, both in its origin and intensity.

The total inventories in the water column are $(2.12 \pm 0.13)\cdot10^{12}$ atoms m$^{-2}$ for $^{236}U$ and $(3.48 \pm 0.18)\cdot10^{12}$ atoms m$^{-2}$ for $^{237}Np$. These results are in agreement with the expected GF inventories in this area estimated from undisturbed soils, of $(2.16 \pm 0.91)\cdot10^{12}$ atoms m$^{-2}$ and $(3.4 \pm 2.1)\cdot10^{12}$ atoms m$^{-2}$ for $^{236}U$ and $^{237}Np$, respectively (i.e. estimations based on the published inventories for $^{239}Pu$ (Hardy et al., 1973) and the $^{236}U/^{239}Pu$ and $^{237}Np/^{239}Pu$ ratios reported in (Elena Chamizo et al., 2015; Kelley et al., 1999)). Figure 4 compares the $^{236}U$ (a) and $^{237}Np$ (b) inventories obtained in this work with the expected GF ones as a function of the latitude. Inventories quantified for other full-depth seawater profiles from different regions are also shown for comparison purposes. It is important to emphasize that, in contrast to soil samples that might keep the inventory of anthropogenic actinides unaltered, the measured inventories in seawater columns are influenced by further water masses redistribution processes and by the physico-chemical behaviour of the elements in the seawater column. However, if local sources are not relevant, a reasonable agreement between both inventories is expected for mostly conservative radionuclides.

Focusing on $^{236}$U, Fig. 4-a shows a clear imbalance between the reported seawater data in the North and the South Hemisphere. Within the North Hemisphere, the only inventories in reasonable agreement with the GF source are those reported for the Japan Sea (Sakaguchi et al., 2012) and the Pacific Ocean (Eigl et al., 2017). North Atlantic (Casacuberta et al., 2014; Castrillejo et al., 2018) and Arctic Oceans results show the influence of liquid effluents released by La Hague NRP into the English channel, and by Sellafield NRP into the Irish Sea. They would have been further transported into those regions through deep water formation and water mass transport processes (Casacuberta et al., 2016). On the other hand, the high inventories for the Mediterranean Sea might be caused by the discharges from Marcoule NRP (France) (Castrillejo et al., 2017; E. Chamizo et al., 2016). In contrast, in the South Atlantic Ocean, the only results are presented in this paper, being similar to the reported data in the Equatorial Pacific. In both cases, the obtained $^{236}$U results are in agreement with the GF source, without apparent influence of other possible local sources.

For $^{237}$Np, Fig. 4-b shows the scarcity of the available information. Apart from this work, there are only two studies providing $^{237}$Np inventories, but both focused on the Mediterranean Sea (Bressac et al., 2017; Castrillejo Iridoy, 2017). As it is shown in Fig. 4-b, all reported inventories are in agreement with GF, and local sources, if existed, would only have a minor effect on those baseline levels.

Therefore, being GF the only relevant source in this area, the studied samples provide an opportunity to get information on the related $^{237}$Np/$^{236}$U atom ratio. However, since this region is affected by upwelling phenomenon, the correct interpretation of $^{236}$U and $^{237}$Np results from surface samples(i.e. corresponding to the MUW) demands a further a discussion about their biogeochemical behaviour in this characteristic environment.

## 5.2 Biogeochemical behaviour of $^{236}$U and $^{237}$Np in the upwelling system

Focusing on the surfaces samples corresponding to MUW, the regional distribution of the $^{236}$U and $^{237}$Np results are depicted in Fig. 5. These surface plots point out to a specific pattern being the lowest concentrations placed at inshore stations. A different view of the data is presented in Fig. 6-a, where $^{236}$U and $^{237}$Np results from surface samples versus distance to the coast for the corresponding sampling stations are plotted. It is evidenced a clear rise of $^{237}$Np concentrations by increasing the distance to the coastline, varying from an average of $5.3\cdot10^6$ atoms kg$^{-1}$ for the shallowest stations (i.e. less than 30 km off the coast, from now on inshore stations) to $8.5\cdot10^6$ atoms kg$^{-1}$ at station 7 (i.e. 170 km off the coast). Such a trend might be caused by Np scavenging effects due to the enhanced presence of particles in shallow waters inshore at the studied site, mostly related to the primary production triggered by the upwelling phenomenon (Louw et al., 2016). Thus, a fraction of the dissolved $^{237}$Np might have been attached to the particles (i.e. organic matter, opal carbonates and crustal material) sinking through the water column and being finally incorporate into the sediment. However, this decrease is not evidenced for $^{236}$U within the instrumental uncertainties. A conclusion that could be drawn is that $^{236}$U has a stronger conservative nature than $^{237}$Np. This hypothesis would be in agreement with the reported partition coefficients for both elements (i.e. $K_d$, the relationship between radionuclide concentrations in suspended particulate matter or bottom sediments and water), which are expected to be higher for Np but still within the range of conservative elements in both cases (IAEA, 2004).

A similar pattern is also observed when plotting $^{236}$U and $^{237}$Np concentrations versus dissolved oxygen (i.e. DO, see Fig. 6-b). The samples with less than 4 mL/L of DO correspond to inshore stations, and have an average $^{237}$Np concentration of $5.32 \cdot 10^6$ atoms kg$^{-1}$ (4% SD), (Table 2). In contrast, the samples with the highest DO (i.e. 4 – 7 mL/L) and corresponding with the stations placed 30 – 170 km off the coastline (from now on offshore stations) present an average $^{237}$Np concentration 25-30% higher (i.e. $7.45 \cdot 10^6$ atoms kg$^{-1}$, 4% SD). Once again, Fig. 6-b does not show a clear pattern in the case of $^{236}$U within uncertainties. However, it is interesting to note that, by calculating the weighted average for inshore and offshore stations, a 10-20% difference appears. This might indicate also an effect of the scavenging in the case of $^{236}$U. The different characteristics of the surface layer for inshore and offshore stations are also reflected in the chlorophyll-a data obtained before and during this sampling campaign (Fig. 7). Furthermore, despite the seasonal variability of the system, previous works developed in this area reported a long-term median concentration of chlorophyll-a and nutrients for inshore stations higher than for offshore ones (Louw et al., 2016). Additionally, it has been reported that, annually, large amounts of nutrients are introduced into the upper layer from water below the thermocline, resulting in elevated chlorophyll-a concentrations in the period of March-April (Louw et al., 2016). Being our sampling camping developed in May, our results might be reflecting a decrease of the dissolved $^{236}$U and $^{237}$Np concentrations as a results of the scavenging effects produced in that intense period.

### 5.3 Characteristic GF $^{237}$Np/$^{236}$U atom ratio in seawater

As it has been previously discussed, the $^{236}$U and specially $^{237}$Np dissolved concentrations in surface samples might be influenced by scavenging effects, decreasing their concentrations notably for the coastal regions and, therefore, the $^{237}$Np/$^{236}$U atom ratios would vary accordingly (Table 2). For the inshore stations, the average $^{237}$Np/$^{236}$U atom ratio is smaller, i.e. 1.25 (6% SD), than the obtained one for offshore stations, i.e. 1.53 (3% SD). Therefore, samples not influenced by this effect have to be considered for the estimation of the GF characteristic ratio. This is the case of two samples from Station 7 collected at 300 m and 600 m corresponding to CW (Table 1), whose weighted average $^{237}$Np/$^{236}$U atom ratio is 1.77 ± 0.20. This value, the very first one directly determined in the same study, can be put in the frame of the available information on that ratio. The expected $^{237}$Np/$^{236}$U ratio for GF can be extrapolated from the reported values for the $^{237}$Np/$^{239}$Pu atom ratio in soils far from local sources in (Kelley et al., 1999), and assuming that the measured $^{236}U/^{239}Pu$ atom ratio for the IAEA reference soil sample Soil-6 (E. Chamizo et al., 2015), collected in Austria, is representative of the GF source in both Hemispheres. Thus, the expected GF $^{237}Np/^{236}U$ atom would be 2.04 ± 0.42 and 1.52 ± 0.40 for the North and the South Hemisphere, respectively. This last value is in agreement with the one measured in this work.

## 6. Summary and conclusions

$^{236}U$ and $^{237}Np$ concentrations have been successfully determined in a set of small volume seawater samples (3-5 L) collected along the Namibian coast, featuring one of the most important upwelling ecosystems in the world. Both radionuclides lay in the $(1\text{-}10)\cdot10^{6}$ atoms $kg^{-1}$ concentration range, achieved thanks to the high sensitivity of the AMS technique. The obtained inventories in the seawater column, $(2.12 \pm 0.13)\cdot10^{12}$ atoms $m^{-2}$ for $^{236}U$ and $(3.48 \pm 0.18)\cdot10^{12}$ atoms $m^{-2}$ for $^{237}Np$, point out to GF as the most relevant source of actinides to the region, overlooked until now in the environmental radioactivity field. Surface samples exhibit a distribution that might be related with the presence of particles in the upwelling region. Inshore samples (i.e. up to 30 km off the coastline) show a clear 25-30% smaller average $^{237}Np$ concentration than offshore ones (i.e. from 30 to 170 km off the coastline), whereas $^{236}U$ concentrations differ roughly by 10-20%. This pattern is similar to the observed ones for other oceanographic parameters such as DO, chlorophyll-a and nutrients. This is the first evidence on a possible enhance particle reactivity of GF Np compared to GF U in a natural marine ecosystem. Moreover, it has been obtained a characteristic $^{237}Np/^{236}U$ atom ratio for the GF source term of 1.77 ± 0.20, figure that is key as for the combined use of $^{236}U$ and $^{237}Np$ in general environmental studies. This work presents the first combined data set of both radionuclides in a marine ecosystem. An achievement has been their measurement from small seawater volumes (3-5 L), which has been possible thanks to the high sensitivity offered by AMS at the CNA, and the reliability of the radiochemical method recently developed in collaboration with the IAEA. More studies are necessary to get new insights into the physicochemical behaviour of $^{237}Np$ and $^{236}U$ in marine environments influenced by different sources, and to improve the accuracy of the $^{237}Np/^{236}U$ GF endmember, in order to set the basis of their use as dual tracer in oceanography.

## Acknowledgments

We are thankful to the Namibian Ministry of Fisheries and Marine Resources (MFMR) for its support, as well as all our colleagues that were involved in the data collection—particularly the staff of the Environment Subdivision. Sampling was done on the RV Mirabilis and captain and crew members from the MFMR are also thanked for their support. This work has been financed from the projects FIS2015-69673-P and PGC2018-094546-B-I00, provided by the Spanish Government (Ministerio de Economía y Competitividad and Ministerio de Ciencia, Innovación y Universidades). This work was partially funded by Fundación Cámara Sevilla through a Grant for Graduate Studies. The IAEA is grateful to the Government of the Principality of Monaco for the support provided to its Environment Laboratories. The authors want to thank to Rafael García-Tenorio and Iolanda Osvath for theguidance on the sampling and analytical strategies, to Isabelle Levy, Oxana Blinova and Frederic Camallonga for the help provided in the management of the samples and to Manuel García León and Simon Jerome for their useful advices in the elaboration of this article.

## References

Beasley, T., Cooper, L.W., Grebmeier, J.M., Aagaard, K., Kelley, J.M., Kilius, L.R., 1998. 237Np/129I atom ratios in the Arctic Ocean: Has 237Np from Western European and Russian fuel reprocessing facilities entered the Arctic Ocean? J. Environ. Radioact. 39, 255–277. https://doi.org/10.1016/S0265-931X(97)00059-3

Bressac, M., Levy, I., Chamizo, E., La Rosa, J.J., Povinec, P.P., Gastaud, J., Oregioni, B., 2017. Temporal evolution of 137 Cs, 237 Np, and 239+240 Pu and estimated vertical 239+240 Pu export in the northwestern Mediterranean Sea. Sci. Total Environ. 595, 178–190. https://doi.org/10.1016/j.scitotenv.2017.03.137

Carr, M.-E., Kearns, E.J., 2003. Production regimes in four Eastern Boundary Current systems. Deep Sea Res. Part II Top. Stud. Oceanogr. 50, 3199–3221. https://doi.org/10.1016/j.dsr2.2003.07.015

Casacuberta, N., Christl, M., Lachner, J., van der Loeff, M.R., Masqué, P., Synal, H.A., 2014. A first transect of 236U in the North Atlantic Ocean. Geochim. Cosmochim. Acta 133, 34–46. https://doi.org/http://dx.doi.org/10.1016/j.gca.2014.02.012

Casacuberta, N., Christl, M., Vockenhuber, C., Wefing, A.-M., Wacker, L., Masqué, P., Synal, H.-A., Rutgers van der Loeff, M., 2018. Tracing the Three Atlantic Branches Entering the Arctic Ocean With $^{129}$ I and $^{236}$ U. J. Geophys. Res. Ocean. 123, 6909–6921. https://doi.org/10.1029/2018JC014168

Casacuberta, N., Masqué, P., Henderson, G., Rutgers van-der-Loeff, M., Bauch, D., Vockenhuber, C., Daraoui, A., Walther, C., Synal, H.-A.A., Christl, M., 2016. First 236U data from the Arctic Ocean and use of 236U/238U and 129I/236U as a new dual tracer. Earth Planet. Sci. Lett. 440, 127–134. https://doi.org/http://dx.doi.org/10.1016/j.epsl.2016.02.020

Castrillejo Iridoy, M., 2017. Sources and distribution of artificial radionuclides in the oceans: from Fukushima to the Mediterranean Sea.

Castrillejo, M., Casacuberta, N., Chirstl, M., Vockenhuber, C., Masqué, P., García-Orellana, J., 2014. Mapping of 236U and 129I in the Mediterranean Sea. Annu. Rep. 2014, Ion Beam Physics, ETH Zurich.

Castrillejo, M., Casacuberta, N., Christl, M., Garcia-Orellana, J., Vockenhuber, C., Synal, H.-A., Masqué, P., 2017. Anthropogenic 236U and 129I in the Mediterranean Sea: First comprehensive distribution and constrain of their sources. Sci. Total Environ. 593–594, 745–759. https://doi.org/10.1016/J.SCITOTENV.2017.03.201

Castrillejo, M., Casacuberta, N., Christl, M., Vockenhuber, C., Synal, H.-A., García-Ibáñez, M.I., Lherminier, P., Sarthou, G., Garcia-Orellana, J., Masqué, P., 2018. Tracing water masses with 129 I and 236 U in the subpolar North Atlantic along the GEOTRACES GA01 section. Biogeosciences 15, 5545–5564. https://doi.org/10.5194/bg-15-5545-2018

Chamizo, E., Christl, M., Fifield, L.K., 2015. Measurement of 236U on the 1MV AMS system at the Centro Nacional de Aceleradores (CNA). Nucl. Instruments Methods Phys. Res. Sect. B Beam Interact. with

Mater. Atoms 358, 45–51. https://doi.org/10.1016/j.nimb.2015.05.008

Chamizo, E., López-Lora, M., 2019. Accelerator mass spectrometry of 236U with He stripping at the Centro Nacional de Aceleradores. Nucl. Instruments Methods Phys. Res. Sect. B Beam Interact. with Mater. Atoms 438, 198–206. https://doi.org/10.1016/j.nimb.2018.04.020

Chamizo, E., López-Lora, M., Bressac, M., Levy, I., Pham, M.K., 2016. Excess of 236U in the northwest Mediterranean Sea. Sci. Total Environ. 565, 767–776. https://doi.org/10.1016/j.scitotenv.2016.04.142

Chamizo, E., López-Lora, M., Bressac, M., Levy, I., Pham, M.K., 2016. Excess of 236 U in the northwest Mediterranean Sea. Sci. Total Environ. 565, 767–776. https://doi.org/10.1016/j.scitotenv.2016.04.142

Chamizo, E., López-Lora, M., Villa, M., Casacuberta, N., López-Gutiérrez, J.M., Pham, M.K., 2015. Analysis of 236U and plutonium isotopes, 239,240Pu, on the 1 MV AMS system at the Centro Nacional de Aceleradores, as a potential tool in oceanography. Nucl. Instruments Methods Phys. Res. Sect. B Beam Interact. with Mater. Atoms 361, 535–540. https://doi.org/10.1016/j.nimb.2015.02.066

Eigl, R., Steier, P., Sakata, K., Sakaguchi, A., 2017. Vertical distribution of 236U in the North Pacific Ocean. J. Environ. Radioact. 169–170, 70–78. https://doi.org/10.1016/J.JENVRAD.2016.12.010

Froehlich, M.B., Akber, A., McNeil, S.D., Tims, S.G., Fifield, L.K., Wallner, A., 2019. Anthropogenic 236U and Pu at remote sites of the South Pacific. J. Environ. Radioact. 205–206, 17–23. https://doi.org/10.1016/J.JENVRAD.2019.05.003

Germain, P., Gandon, R., Masson, M., Guéguéniat, P., 1987. Experimental studies of the transfer of neptunium from sea water to sediments and organisms (annelids and molluscs). J. Environ. Radioact. 5, 37–55. https://doi.org/10.1016/0265-931X(87)90043-9

Hallstadius, L., Aarkrog, A., Dahlgaard, H., Holm, E., Boelskifte, S., Duniec, S., Persson, B., 1986. Plutonium and americium in arctic waters, the North Sea and Scottish and Irish coastal zones. J. Environ. Radioact. 4, 11–30. https://doi.org/10.1016/0265-931X(86)90018-4

Hardy, E.P., Krey, P.W., Volchok, H.L., 1973. Global Inventory and Distribution of Fallout Plutonium. Nature 241, 444–445. https://doi.org/10.1038/241444a0

Hutchings, L., van der Lingen, C.D., Shannon, L.J., Crawford, R.J.M., Verheye, H.M.S., Bartholomae, C.H., van der Plas, A.K., Louw, D., Kreiner, A., Ostrowski, M., Fidel, Q., Barlow, R.G., Lamont, T., Coetzee, J., Shillington, F., Veitch, J., Currie, J.C., Monteiro, P.M.S., 2009. The Benguela Current: An ecosystem of four components. Prog. Oceanogr. 83, 15–32. https://doi.org/10.1016/J.POCEAN.2009.07.046

IAEA, 2004. Sediment Distribution Coefficients and Concentration Factors for Biota in the Marine Environment. Vienna.

International Atomic Energy Agency., 1997. Planning and operation of low level waste disposal facilities : proceedings of an International Symposium on Experience in the Planning and Operation of Low Level Waste Disposal Facilities organized by the International Atomic Energy Agency and held in Vienna, 17-21 June 1996. International Atomic Energy Agency.

Kelley, J.M., Bond, L.A., Beasley, T.M., 1999. Global distribution of Pu isotopes and 237Np. Sci. Total Environ. 237–238, 483–500. https://doi.org/http://dx.doi.org/10.1016/S0048-9697(99)00160-6

Lindahl, P., Roos, P., Holm, E., Dahlgaard, H., 2005. Studies of Np and Pu in the marine environment of SwedisheDanish waters and the North Atlantic Ocean. J. Environ. Radioact. 82, 285–301. https://doi.org/10.1016/j.jenvrad.2005.01.011

López-Lora, M., Chamizo, E., 2019. Accelerator Mass Spectrometry of 237Np, 239Pu and 240Pu for environmental studies at the Centro Nacional de Aceleradores. Nucl. Instruments Methods Phys. Res. Sect. B Beam Interact. with Mater. Atoms 455, 39–51. https://doi.org/10.1016/j.nimb.2019.06.018

López-Lora, M., Levy, I., Chamizo, E., 2019. Simple and fast method for the analysis of 236U, 237Np, 239Pu and 240Pu from seawater samples by Accelerator Mass Spectrometry. Talanta 200, 22–30. https://doi.org/10.1016/j.talanta.2019.03.036

Louw, D.C., van der Plas, A.K., Mohrholz, V., Wasmund, N., Junker, T., Eggert, A., 2016. Seasonal and interannual phytoplankton dynamics and forcing mechanisms in the Northern Benguela upwelling system. J. Mar. Syst. 157, 124–134. https://doi.org/10.1016/j.jmarsys.2016.01.009

McCubbin, D., Leonard, K.S., 1997. Laboratory studies to investigate short-term oxidation and sorption behaviour of neptunium in artificial and natural seawater solutions. Mar. Chem. 56, 107–121. https://doi.org/10.1016/S0304-4203(96)00085-0

Pentreath, R.J., Harvey, B.R., 1981. The presence of 237Np in the Irish Sea. Mar. Ecol. Prog. Ser. 6, 243–247.

Peterson, R.G., Stramma, L., 1991. Upper-level circulation in the South Atlantic Ocean. Prog. Oceanogr. 26, 1–73. https://doi.org/10.1016/0079-6611(91)90006-8

Sakaguchi, A., Kadokura, A., Steier, P., Takahashi, Y., Shizuma, K., Hoshi, M., Nakakuki, T., Yamamoto, M., 2012. Uranium-236 as a new oceanic tracer: A first depth profile in the Japan Sea and comparison with caesium-137. Earth Planet. Sci. Lett. 333–334, 165–170. https://doi.org/10.1016/j.epsl.2012.04.004

Shannon, L.V., 1985. The Benguela Ecosystem, Part 1. Evolution of the Benguela, physical features and processes., in: Oceanography and Marine Biology: An Annual Review 23. pp. 105–182.

Steier, P., Bichler, M., Keith Fifield, L., Golser, R., Kutschera, W., Priller, A., Quinto, F., Richter, S., Srncik, M., Terrasi, P., Wacker, L., Wallner, A., Wallner, G., Wilcken, K.M., Maria Wild, E., 2008. Natural and anthropogenic 236U in environmental samples. Nucl. Instruments Methods Phys. Res. Sect. B Beam Interact. with Mater. Atoms 266, 2246–2250. https://doi.org/10.1016/j.nimb.2008.03.002

Thakur, P., Mulholland, G.P., 2012. Determination of 237Np in environmental and nuclear samples: A review of the analytical method. Appl. Radiat. Isot. 70, 1747–1778. https://doi.org/10.1016/j.apradiso.2012.02.115

UNSCEAR, 2008. Sources and effect of ionizing radiation. UNSCEAR 2008 Report Vol. I.

Villa-Alfageme, M., Chamizo, E., Kenna, T.C., López-Lora, M., Casacuberta, N., Chang, C., Masqué, P., Christl, M., 2019. Distribution of 236U in the U.S. GEOTRACES Eastern Pacific Zonal Transect and its use as a water mass tracer. Chem. Geol. 517, 44–57. https://doi.org/10.1016/j.chemgeo.2019.04.003

Vorrath, M.E., Lahajnar, N., Fischer, G., Libuku, V.M., Schmidt, M., Emeis, K.C., 2018. Spatiotemporal variation of vertical particle fluxes and modelled chlorophyll a standing stocks in the Benguela Upwelling System. J. Mar. Syst. 180, 59–75. https://doi.org/10.1016/j.jmarsys.2017.12.002

Winkler, S.R., Steier, P., Carilli, J., 2012. Bomb fall-out 236U as a global oceanic tracer using an annually resolved coral core. Earth Planet. Sci. Lett. 359–360, 124–130. https://doi.org/10.1016/J.EPSL.2012.10.004

## Tables

**Table 1-** Obtained $^{236}U/^{238}U$ ratios and $^{237}Np$ concentrations for the different water masses identified according to the T-S plots (Fig. 3). Reported $^{236}U/^{238}U$ atom ratios for AAIW and NADW in the North Atlantic Ocean (Casacuberta et al., 2014) are also shown for comparison purposes.

| Samples | Water masses | Measured values | | Reported values |
|---|---|---|---|---|
| | | **$^{236}U/^{238}U$ (atom ratio)** | **$^{237}Np$ ($10^6$ at $kg^{-1}$)** | **$^{236}U/^{238}U$ ( atom ratio)** |
| Surface samples (average) | MUW | $(5.53 \pm 0.11)\cdot10^{-10}$ | 6.24 ± 0.15 | - |
| 300 m at St. 7 | CW | $(3.82 \pm 0.31)\cdot10^{-10}$ | 5.56 ± 0.42 | - |
| 600 m at St. 7 | CW | $(7.1 \pm 1.8)\cdot10^{-11}$ | 1.08 ± 0.19 | - |
| 1250 m at St.7 | AAIW<br>NADW | $(2.2 \pm 1.7)\cdot10^{-11}$ | < 1 | AAIW: $(2\text{-}4)\cdot10^{-10}$<br>NADW: $<1.5\cdot10^{-10}$<br>(Casacuberta et al., 2014) |

**Table 2-** Average $^{236}U$ and $^{237}Np$ results for surface samples considering two different ranges of dissolved oxygen concentrations: corresponding to distances to de coastline of 0 – 30 km (i.e. inshore stations) and 30 – 70 km (i.e. offshore stations). Average weighted values and standard deviations (SD) have been weighted by the corresponding uncertainties.

| Coast distance (km) | Oxygen (mL/L) | $^{236}U/^{238}U$ | | $^{236}U$ | | $^{237}Np$ | | $^{237}Np/^{236}U$ | |
|---|---|---|---|---|---|---|---|---|---|
| | | **atom ratio** | **SD (%)** | **$10^6$ at $kg^{-1}$** | **SD (%)** | **$10^6$ at $kg^{-1}$** | **SD (%)** | **atom ratio** | **SD (%)** |
| 0 – 30 | 0 – 4 | 5.25 | 3% | 4.30 | 3% | 5.32 | 4% | 1.25 | 6% |
| 30 – 170 | 4 – 7 | 5.86 | 3% | 4.94 | 3% | 7.45 | 2% | 1.53 | 3% |

## Figures

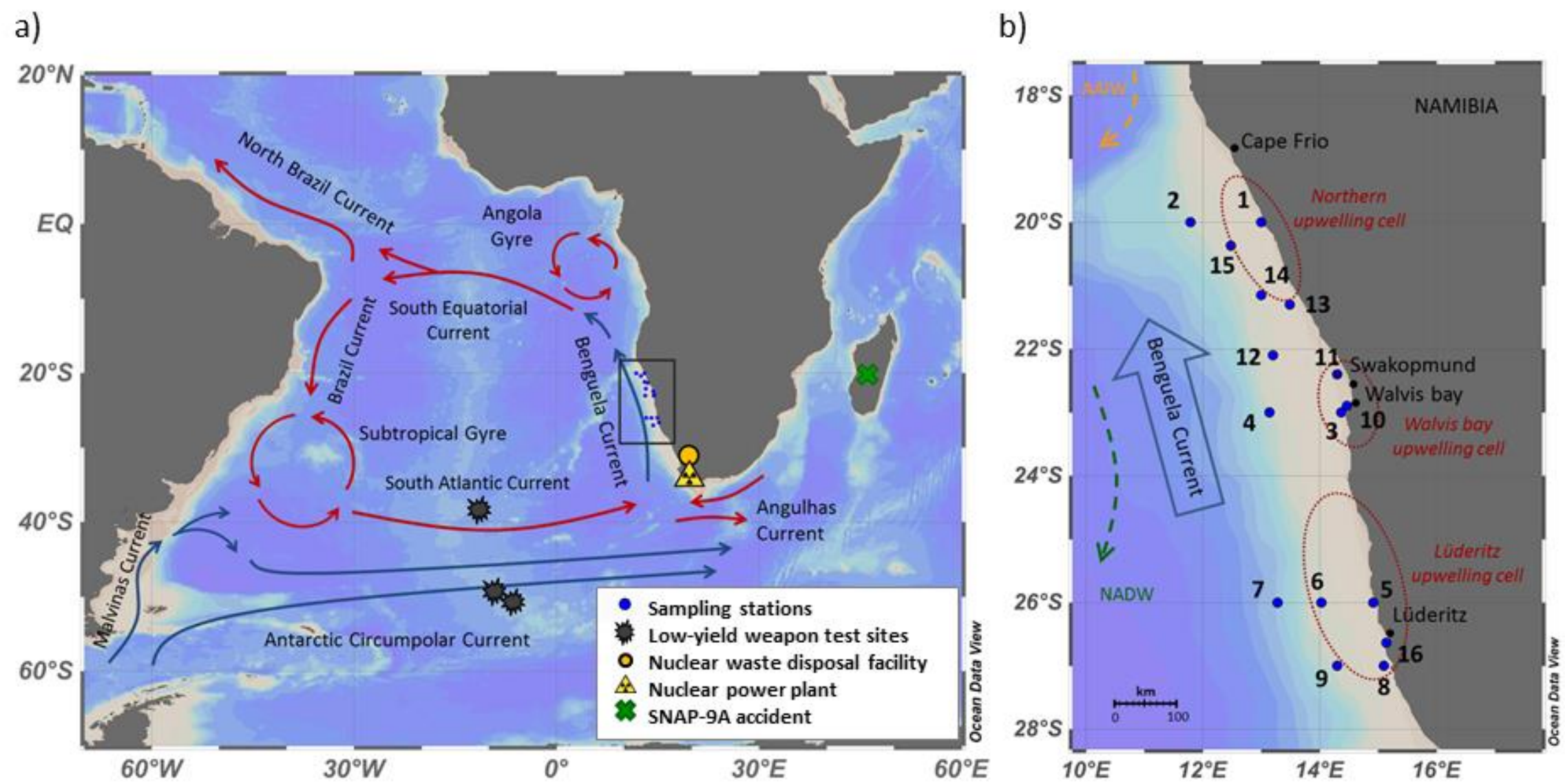

**Fig. 1- (a)** Map showing the location of the studied area and the most important surface circulation currents in the South Atlantic Ocean (red arrows indicate generally warmer water currents and blue arrows indicate generally cooler waters ) (Stramma and England, 1999). Potential $^{236}U$ and $^{237}Np$ local sources to this region are also shown. (b) The sampling stations are depicted together with the main local currents including the deep currents associated to Antarctic Intermediate Water (AAIW) and North Atlantic Deep Water (NADW) (Stramma and England, 1999). The most important upwelling cells are also indicated.

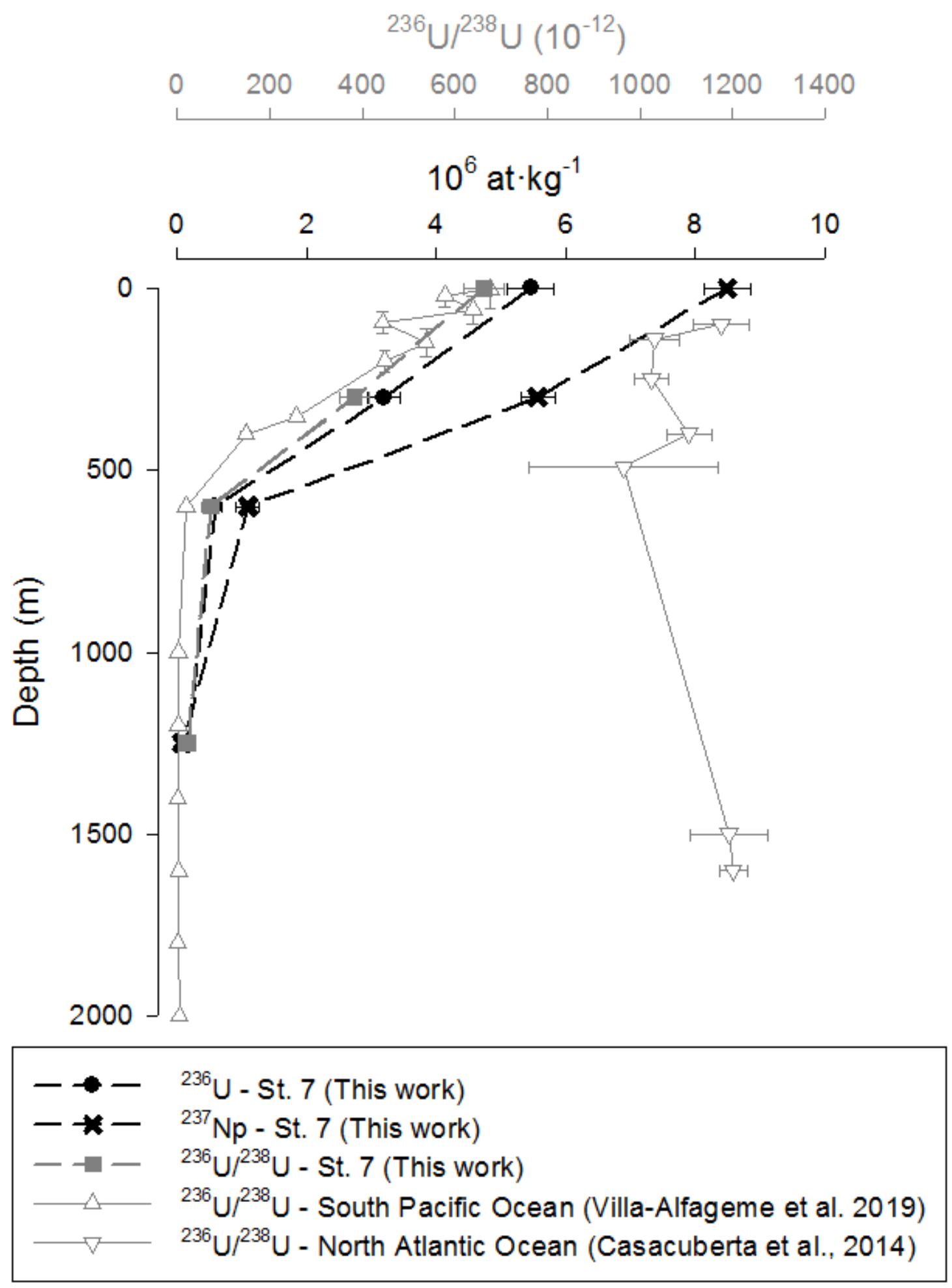

**Fig. 2-** $^{236}U$ and $^{237}Np$ results for the depth profile at Station 7 (see Fig. 1). $^{236}U/^{238}U$ atom ratios reported from the North Atlantic Ocean (St. 2 at 64°N, (Casacuberta et al., 2014)) and other from the South Pacific Ocean (St.18 at 15°S, (Villa-Alfageme et al., 2019)) are also shown for comparison purposes.

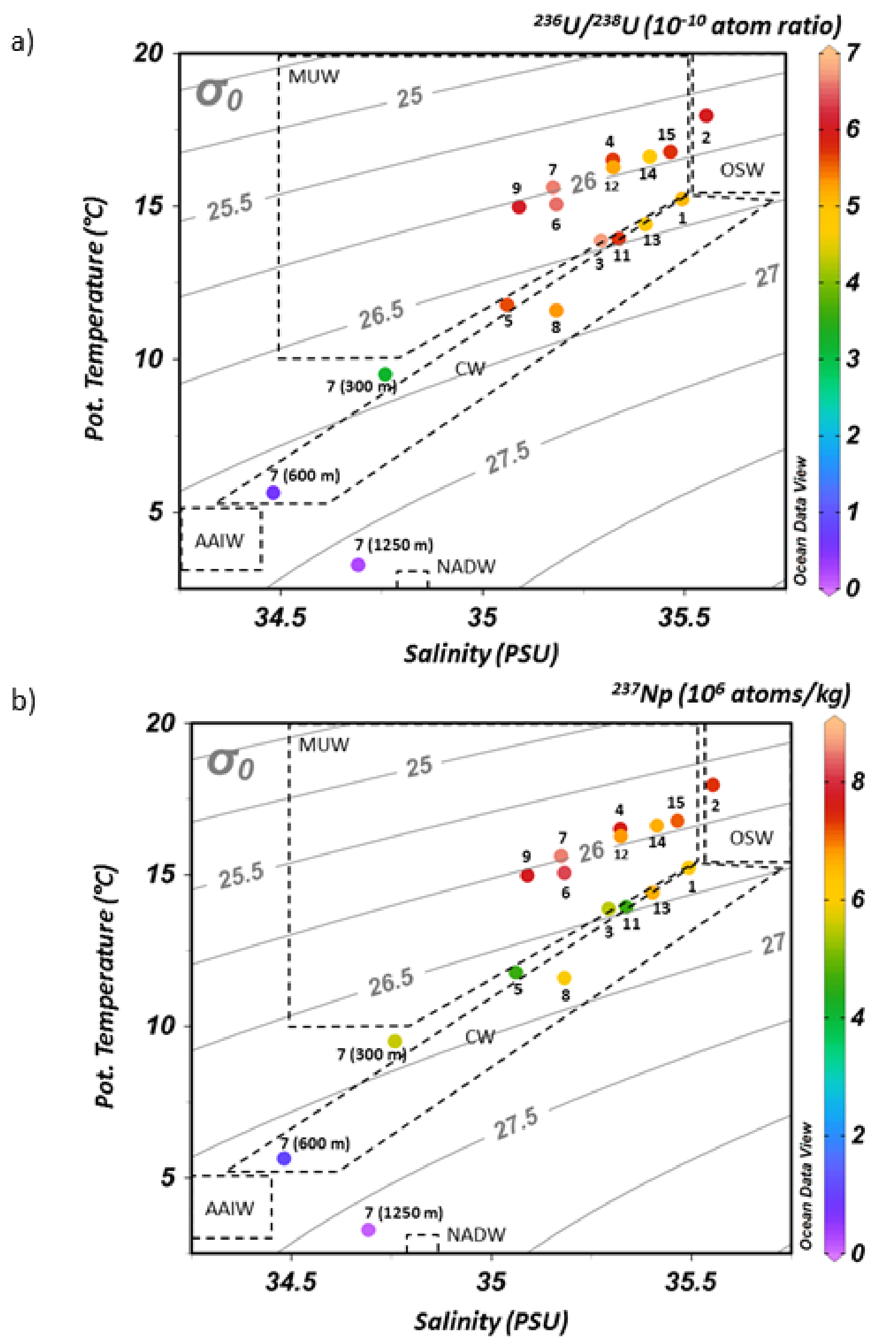

**Fig. 3-** T-S (potential temperature vs salinity) plots with $^{236}U/^{238}U$ atom ratios (a) and $^{237}Np$ concentrations (b). Isopycnals ($\sigma_0$) and station numbers indicated in the Fig. 1 are also shown. Having shallow depths, Stations 10 and 16 were sampled with an auxiliary boat and no T-S data exists, so they have not been included in the plot (Louw, 2014). Water masses are indicated following the T-S ranges given by (Rae, 2005) . The labelled water masses are: Modified Upwelling Waters (MUW), Oceanic Surface Water (OSW), Central Waters (CW), Antarctic Intermediate Waters (AAIW) and North Atlantic Depth Waters (NADW).

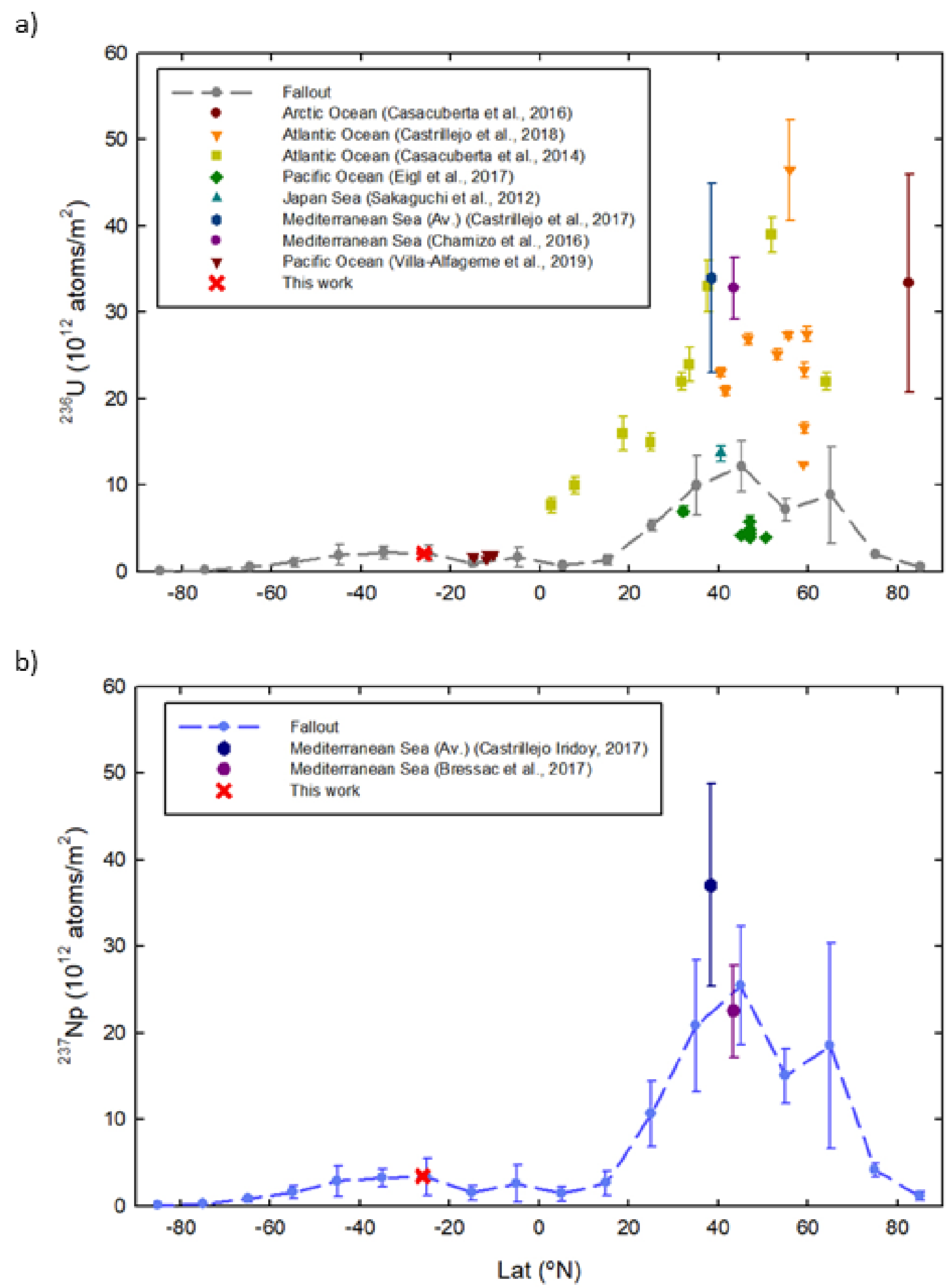

**Fig. 4-** $^{236}$U (a) and $^{237}$Np (b) inventories obtained in this work the full-depth profile at Station 7, together with the estimated GF inventories from soil studies as a function of the latitude (i.e. estimations from $^{239}$Pu results in (Hardy et al., 1973) and the $^{236}$U/$^{239}$Pu and $^{237}$Np/$^{239}$Pu ratios reported in (Elena Chamizo et al., 2015; Kelley et al., 1999). Reported seawater column inventories from previous works are also shown for comparison purposes. The average $^{236}$U and $^{237}$Np values are shown in the graphs in the case of the Mediterranean Sea results reported by (Castrillejo et al., 2017) and (Castrillejo Iridoy, 2017), respectively.

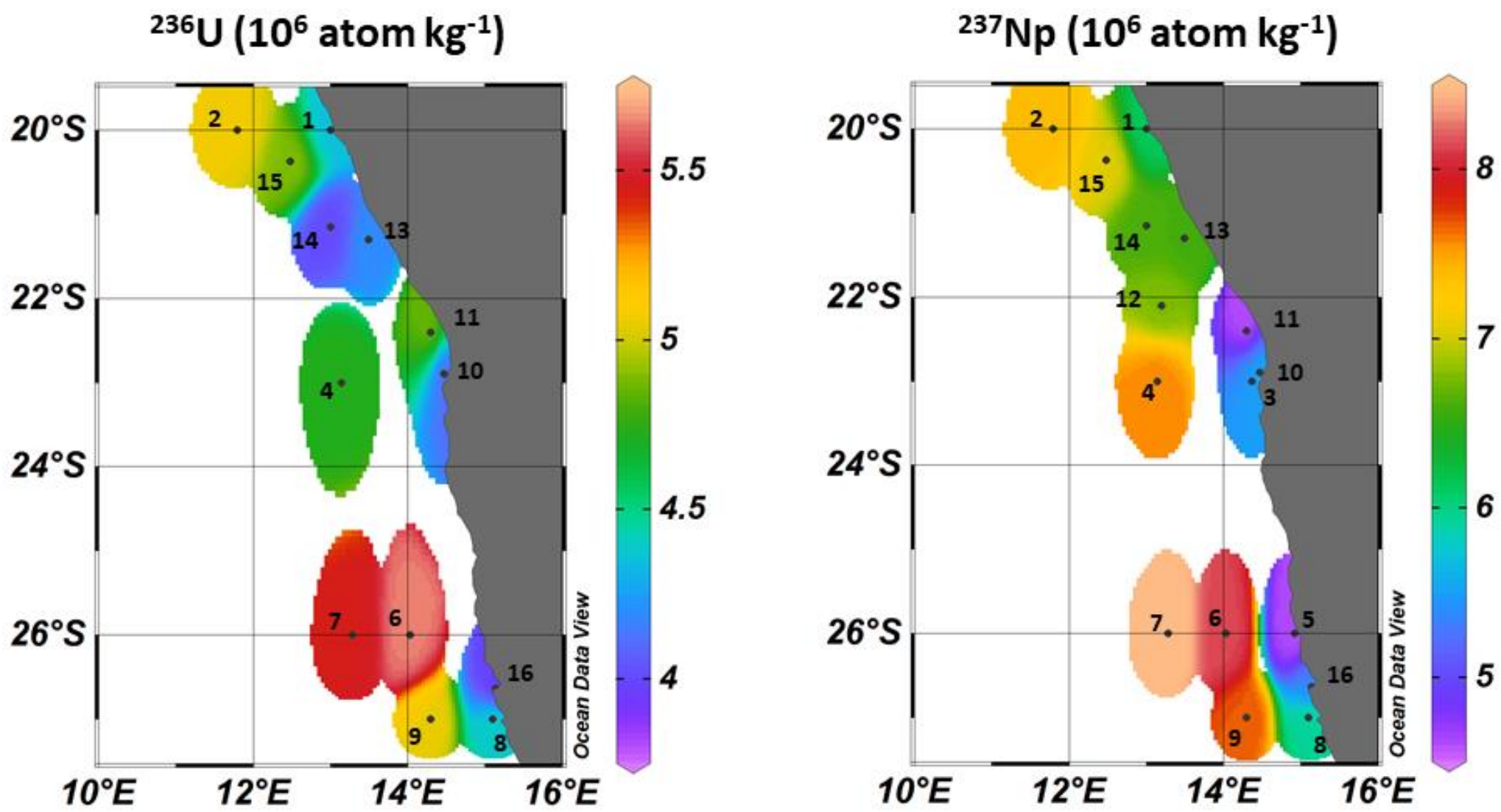

**Fig. 5-** Surface plots of the $^{236}$U and $^{237}$Np concentrations for the studied surface seawater samples. Only results with relative uncertainties of 10% or lower are shown. Station numbers are also indicated.

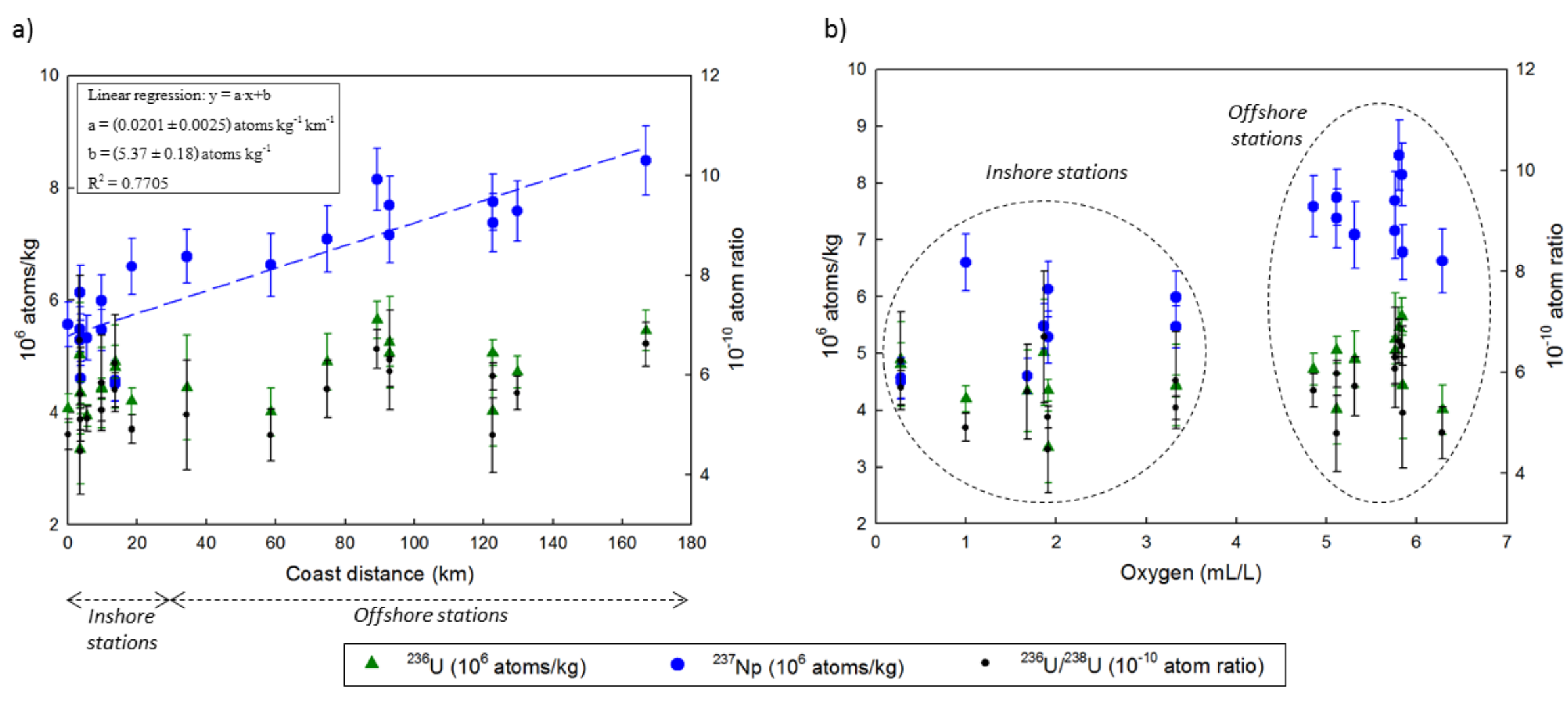

**Fig. 6-** Plots showing the $^{236}$U and $^{237}$Np concentrations and the $^{236}$U/$^{238}$U atom ratios in surface samples versus the distance to the coastline of the corresponding stations (a) and the dissolved oxygen in the samples (b). In the case of $^{237}$Np results in the first plot, the linear fit is also indicated. Moreover, the corresponding ranges for inshore and offshore stations are also shown.

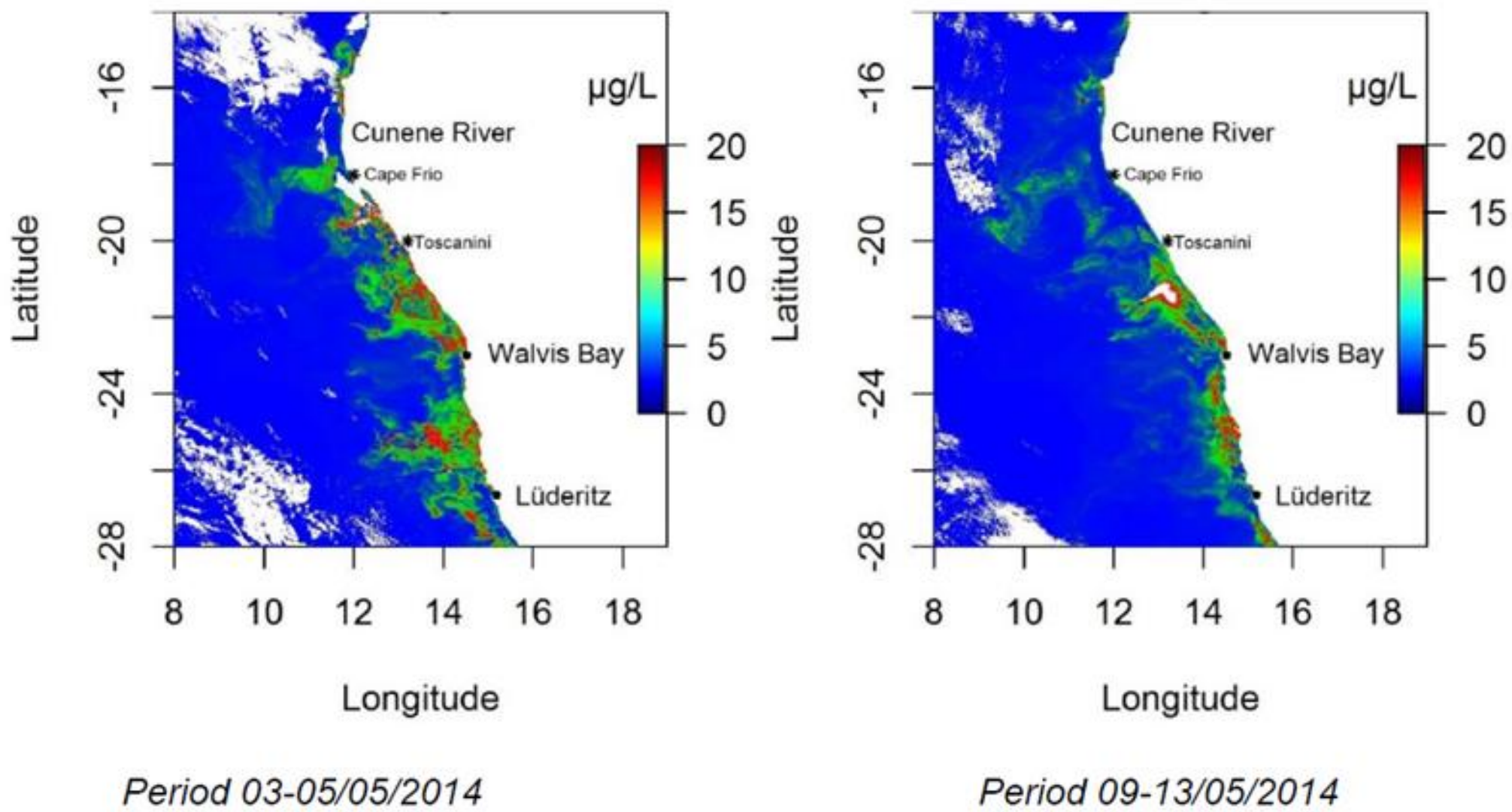

**Fig. 7-** Average remote sensing chlorophyll-a concentrations off Namibia before (left) and during (right) the sampling campaign, showing the dense phytoplankton blooms close inshore [data was obtained from NASA] (Louw, 2014).

# Supplemental information

## $^{238}$U results

Figure S1 compares the $^{238}$U concentrations in this work with the predicted ones from the salinity data using the semi-empirical formula given in (Pates and Muir, 2007). The empirical data differ from the salinity-based results in some cases. Although these discrepancies appear by chance in the most of the CNA samples collected at inshore stations (i.e. stations less than 30 km away from the coastline and with dissolved oxygen values of 0 – 4 mL/L), the corresponding IAEA duplicates are in agreement with the expected values. Thus, this might indicate a problem of homogeneity of the first batch of samples being acidified in the laboratory in contrast with the IAEA samples which were acidified on board. Having been collected without an initial acidification, a fraction of the dissolved $^{238}$U would have been attached to the walls of the container, and a longer acidification period in the laboratory would have been necessary. Nevertheless, this effect has not been observed in the case of the anthropogenic radionuclides, taking into account their corresponding uncertainties.

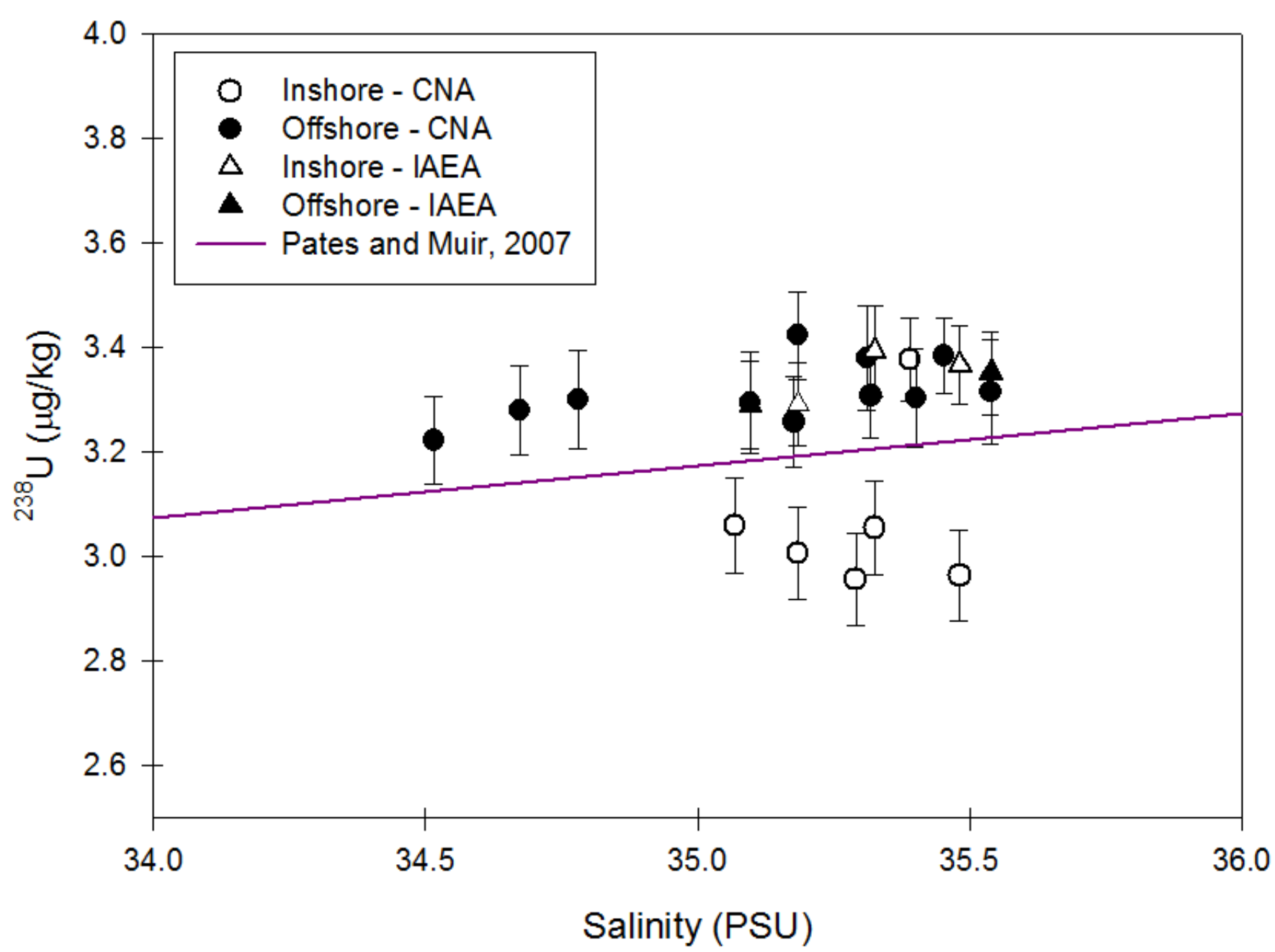

***Fig. S1-*** *$^{238}$U concentrations for all the samples analysed in this work as a function of the salinity. Inshore stations (empty symbols) are the ones placed at less than 30 km to the coastline bottom and offshore stations (fill symbols) the ones between 30 km and 170 km to the coast. Aliquots from the IAEA repository have been indicated with triangles (see Section 3.2 for details). The salinity based $^{238}$U concentrations for the 34-36 PSU range, obtained using the formula $^{238}U[\mu g/kg] = (0.0931 \pm 0.0016)\cdot S$ given in (Pates and Muir, 2007) are also shown.*

## Data availability

The data that support the findings of this study are openly available at the following URL:

**https://github.com/AMS-CNA/NamibiaSW**